# Quantum circuits design for evaluating transcendental functions based on a function-value binary expansion method

Shengbin Wang[1], Zhimin Wang[1,3], Wendong Li[1], Lixin Fan[1], Guolong Cui[1], Zhiqiang Wei[2] and Yongjian Gu[1,3]

[1] Department of Physics, College of Information Science and Engineering, Ocean University of China, Qingdao 266100, China

[2] Department of Computer Science and Technology, College of Information Science and Engineering, Ocean University of China, Qingdao 266100, China

E-mail: wangzhimin@ouc.edu.cn and guyj@ouc.edu.cn

**Abstract**

Quantum arithmetic in the computational basis constitutes the fundamental component of many circuit-based quantum algorithms. There exist a lot of studies about reversible implementations of algebraic functions, while research on the higher-level transcendental functions is scant. We propose to evaluate the transcendental functions based on a novel methodology, which is called qFBE (quantum Function-value Binary Expansion) method. This method transforms the evaluation of transcendental functions to the computation of algebraic functions in a simple recursive way. We present the quantum circuits for solving the logarithmic, exponential, trigonometric and inverse trigonometric functions based on the qFBE method. The efficiency of the circuits is demonstrated on a quantum virtual computing system installed on the Sunway TaihuLight supercomputer. The qFBE method provides a unified and programmed solution for the evaluation of transcendental functions, and it will be an important building block for many quantum algorithms.

## 1. Introduction

After decades of efforts, we are beginning to realize the promise of quantum computing that quantum computers have capabilities surpassing what classical computers can do [1,2]. When we speak of quantum speedup, we typically consider asymptotic scaling of complexity measures of quantum algorithm with problem size [3]. Now from an implementation-of-algorithm point of view, ongoing work is desired to provide more detailed cost estimates and practicality analysis of various quantum algorithms. To do this, quantum arithmetic would be the always-encountered obstacle.

Arithmetic in the computational basis constitutes the fundamental component of many circuit-based quantum algorithms. The most famous example is the modular exponentiation in Shor's factoring algorithm [4]. The trigonometric and inverse trigonometric functions are used in Grover's state preparation algorithm [5], the HHL algorithm for linear equations [6], and quantum analog-digital conversion algorithm [7], *etc*. The basis arithmetic would divide into evaluation of algebraic and transcendental functions, of which the former includes addition, multiplication, inverse and square root, *etc.*, and the latter includes the logarithmic, exponential, trigonometric and inverse trigonometric functions, *etc*.

---

[3]Author to whom any correspondence should be addressed.

For the algebraic functions, there exists a vast amount of literature which provide quantum circuits for addition [8-11], multiplication [8,12-15], reciprocal [16,17,21], and square root [18-21]. However, studies about quantum arithmetic of transcendental functions are sparse. Bhaskar et al. [19] make use of Newton iteration and repeated square root method to evaluate logarithmic function. Cao et al. [16] propose to evaluate sin($x$) based on Taylor expansion and repeated squaring method, and then compute arcsin($x$) by several invocations of sin($x$). Recently, Häner et al. [20] develop a method of piecewise polynomial approximation to evaluate sin($x$), arcsin($x$), exp(-$x$), exp(-$x^2$) and tanh($x$) with different choices of parameters. By and large, the methodologies used by these studies are single, which are mainly polynomial approximations. And the cost of the resulting circuits is still quite high, or the error propagation through the circuit is hard to control.

In the present work, we propose a new way of evaluating transcendental functions based on a completely different methodology. We refer to this method as qFBE (abbreviation for quantum Function-value Binary Expansion) method. Roughly speaking, the function-value binary expansion method transforms the evaluation of transcendental functions into evaluation of algebraic functions, and outputs the binary string of the function value digit-by-digit in a simple recursive way [22]. So qFBE can make good use of existing algorithms for algebraic functions, and make circuits design more compact and modular. It is intriguing to mention that the methodology of qFBE is very similar with that of the classical CORDIC (COordinate Rotation DIgital Computer) arithmetic [23]. In classical computing, CORDIC has been demonstrated as a better choice for scientific calculator applications because of its potential for unified solution for a large set of computational tasks and its simplicity of hardware implementation [24]. So qFBE may represent a more suitable and efficient way for implementing transcendental functions in quantum computer.

We have used qFBE method to evaluate arc cotangent function to implement the controlled rotation operation in the HHL algorithm [21]. The present paper will discuss qFBE in details. The structure of the paper is as follows. In Section 2, we describe the general idea of classical function-value binary expansion method as well as the further improvements we made for broader applications. In Section 3, we provide detailed designs of quantum circuits for various transcendental functions based on the qFBE method. Section 4 shows the potential that qFBE can be extended to the qudit system. In Section 5, we demonstrate the quantum circuits on a quantum virtual system on the Sunway TaihuLight supercomputer. Finally, conclusions are given in Section 6.

## 2. Classical function-value binary expansion method

The phenomenon that a function-value of arctan($x$) can be expanded into a binary form was first discovered by Simon Plouffe using a numerical method [25]. Let $x \geq 0$, the function value of arctan($x$) can be expressed as

$$\frac{\arctan(x)}{\pi} = \sum_{n \geq 0,\ a_n < 0} \frac{1}{2^{n+1}}.$$

$$\text{Let } a_0 = x,\ a_{n+1} = \frac{2a_n}{1-a_n^2}\ (a_{n+1} = -\infty\ \text{if}\ a_n = \pm 1).$$

(1)

Then Borwein and Girgensohn explored the underlying principle of the identity, and utilized it to get similar recursions and binary expansions for other functions. They found that the problem was actually to deal with a system of functional equations which was named system (S) [22]. Set an interval $I \subseteq \mathbb{R}$, system (S) represents the following functional equations for an unknown function $f: I \to [0, 1]$,

$$\begin{cases} 2f(x) = f(r_0(x)) & \text{if } x \in D_0 \\ 2f(x) - 1 = f(r_1(x)) & \text{if } x \in D_1 \end{cases},$$

(2)

where $D_0$ and $D_1$ are subintervals of $I$ with $D_0 \cup D_1 = I$, $D_0 \cap D_1 = \emptyset$; $r_0$ and $r_1$ are functions defined as $r_0: D_0 \to I,\ r_1: D_1 \to I$. Stating the functional equations in this way implies that for each solution of $f(x)$ the following is true,

$$\begin{cases} x \in D_0 \Rightarrow f(x) \in [0, 1/2] \\ x \in D_1 \Rightarrow f(x) \in [1/2, 1] \end{cases}.$$

(3)

This is an important feature of the functions belonging to system (S), i.e. satisfying Eq. (2). According to Eq. (3), a function value $f(x)$ being greater or less than 1/2 can be determined by the subinterval the point $x$ is in. Furthermore, Eq. (3) indicates the dividing point between $D_0$ and $D_1$, which is actually the point that $f(x) = 1/2$.

Based on Eqs. (2) and (3), we can obtain the following fact that for any functions being members of system (S), the function value can be expanded in a binary form as

$$\text{Let } a_0 = x,\ a_{n+1} = \begin{cases} r_0(a_n) & \text{if } a_n \in D_0 \\ r_1(a_n) & \text{if } a_n \in D_1 \end{cases},$$

$$\text{Then } f(x) = \sum_{n \geq 0,\ a_n \in D_1} \frac{1}{2^{n+1}}.$$

(4)

This is just the function-value binary expansion method. Fig. 1 may be helpful to understand the recursion. Let's see it step by step. (You should observe that the following procedure is just like that of computing the binary string of a normal decimal.)

Step 1, namely $n = 0$. Assume point $x_0$ in subinterval $D_0$, so $f(x_0)$ is less than 1/2. Then we should multiply $f(x_0)$ by 2, and the corresponding abscissa is $r_0(x_0)$ belonging to $D_1$, i.e., $2f(x_0) = f(r_0(x_0))$, as shown by the red arrow in Fig. 1.

Step 2, namely $n = 1$. Now $r_0(x_0)$ in subinterval $D_1$, so $2f(x_0)$ is greater than 1/2. That is, the binary string of $f(x_0)$ have the term of $1/2^2$. Then we should do the operation of $2(2f(x_0))-1$, and the corresponding abscissa is $r_1(r_0(x_0))$ belonging to $D_1$, i.e., $2(f(r_0(x_0)))-1 = f(r_1(r_0(x_0)))$.

Step 3, namely $n = 2$ (which is not shown in Fig. 1). The x-coordinate is in subinterval $D_1$, so $2(2f(x_0))-1$ is greater than $1/2$. That is, the binary string of $f(x_0)$ have the term of $1/2^3$. Then we should do the operation of $2(2(2f(x_0))-1)-1$, and the corresponding abscissa is $r_1(r_1(r_0(x_0)))$……

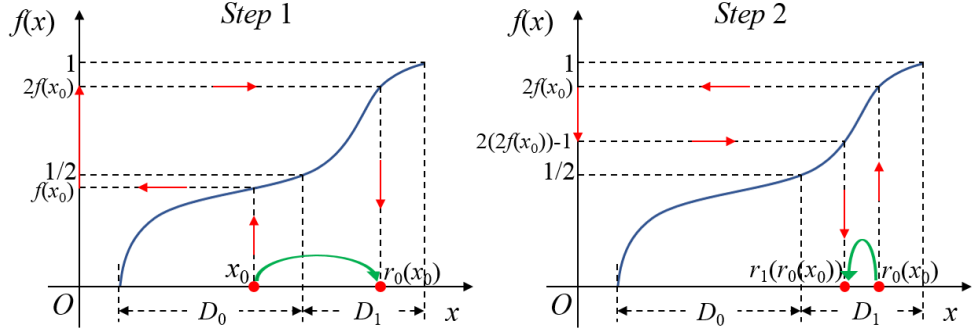

Fig.1. Illustration of the classical function-value binary expansion method.

The recursion can work due to the fact that there exit proper $r_0(x)$ and $r_1(x)$ for the objective function $f(x)$. In fact, the challenge is to find proper $r_0(x)$ and $r_1(x)$ which should be simpler than $f(x)$, e.g., $f$ is transcendental and $r_0$ and $r_1$ are algebraic. In principle, $r_0(x)$ and $r_1(x)$ can be obtained by solving the two equations in Eq. (2). Taking the logarithmic function as an example, $2\ln(x) = \ln(x^2) = \ln(r_0(x))$, then $r_0(x) = x^2$. However in general, it may need certain tricks to find proper $r_0$ and $r_1$ for an objective function.

The frequently-used logarithmic and inverse trigonometric functions are members of system (S) listed in Group 1 as follows,

**Group 1**: $\log_2(x)$ & $\ln(x)$, $\arccos(x)$ & $\arcsin(x)$, $\text{arccot}(x)$ & $\arctan(x)$.

But the exponential and trigonometric functions are not members of system (S) listed in Group 2 as follows,

**Group 2**: $2^x$ & $e^x$, $\cos(x)$ & $\sin(x)$, $\cot(x)$ & $\tan(x)$.

Note that functions in Group 2 are actually the inverse functions of those in Group 1. From this point, we find that the function-value binary expansion method can be inversed to evaluate the corresponding inverse functions in the following way,

$$\text{Set } x = (0.v_{n-1}v_{n-2}\cdots v_1v_0), v_i \in \{0,1\},$$

$$\text{Let } a_0 = const, a_{i+1} = \begin{cases} r_0^{-1}(a_i) & \text{if } v_i = 0 \\ r_1^{-1}(a_i) & \text{if } v_i = 1 \end{cases}, \quad (5)$$

$$\text{Then } f(x) = a_{n+1}.$$

Comparing this equation with Eq. (4), the ways of evaluating functions are different, or in some sense opposite. In Eq. (4), the function value is evaluated digit-by-digit in a recursive way, while in Eq. (5), the function value is approximated stage-by-stage in an iterative way.

In the following subsequent sections, we will develop the corresponding quantum version of the function-value binary expansion method and present quantum circuits for implementing the transcendental functions in both Group 1 and Group 2.

## 3. Quantum circuits for implementing transcendental functions

In this section, we will first show the detailed circuits design for evaluating the functions in Group 1 by the quantum function-value binary expansion method (*i.e.*, qFBE), and then show the circuits for the functions in Group 2 by the inversed quantum function-value binary expansion method (*i.e.*, inversed qFBE).

3.1 Quantum circuits for functions in Group 1
3.1.1 Logarithmic function
We will show the algorithm and circuit for the base-two logarithmic function. Other kinds of logarithmic functions can be evaluated easily using the base changing formula. The basic elements used to expand values of base-two logarithmic function into binary form are listed as follows,

$$f(x) = \log_2 x, \; x \in I = [1, 2),$$
$$D_0 = [1, \sqrt{2}), \; D_1 = [\sqrt{2}, 2), \tag{6}$$
$$r_0(x) = x^2, \quad r_1(x) = x^2/2.$$

The domain $I$ is chosen to make the function satisfy the definition of $f : I \to [0, 1]$.

Substitute them into Eq. (4), and we can evaluate the logarithmic function by computing square recursively. The final output is written as

$$f(x) = (0.w_0 w_1 \cdots w_{n-1})_2, w_i \in \{0,1\}. \tag{7}$$

Let us take the evaluation of $f(1.5)$ as the example to illustrate the recursive process. (The value of $f(1.5)$ is about 0.585 which is 0.100101 up to six digits.) First, let $a_0$ be equal to the input value 1.5. Then it goes as follows.
  1st. Since $a_0 \in D_1$, we have $w_0 = 1$, and $a_1 = r_1(a_0) = 1.125$.
  2nd. Since $a_1 \in D_0$, we have $w_1 = 0$, and $a_2 = r_0(a_1) = 1.265625$.
  3rd. Since $a_2 \in D_0$, we have $w_2 = 0$, and $a_3 = r_0(a_2) \approx 1.601807$.
  4th. Since $a_3 \in D_1$, we have $w_3 = 1$, and $a_4 = r_1(a_3) \approx 1.282893\ldots$

As can be seen from step 3, the intermediate value of $a_i$ brings truncation error, which would accumulate and finally cause flips of some bit, say $w_j$. However, if the precision of intermediate value $a_i$ is high enough, it can be guaranteed that each digit of the output $w$ is accurate and the solution error comes only from the truncation of binary bits. Specific analysis of the error accumulation for all the functions in Groups 1 and 2 are discussed in the appendix.

In quantum computing, the input can be in superposition states, so some of the input values may be not in the domain *I*. In order to address this problem, we expand the function domain in the following way. First, we shift the input values to the left or right to make the minimum one be involved in the interval $I = [1, 2)$. The input $x$ turns to $y = 2^{\pm p} x$ after shifting, where $p$ represents the shifted bits. Note that since all calculations are implemented in fixed precision arithmetic, this shift operation can be performed easily by just keeping in track the position of the binary point. Suppose now the magnitude of the difference between max and min values is $2^q - 1 \leq \Delta y < 2^{q+1} - 1$, then Eq. (6) can be expanded to the following equations,

$$f(y) = (\log_2 y)/(q+1), \ y \in I = [1, 2^{q+1}),$$
$$D_0 = [1, 2^{(q+1)/2}), \ D_1 = [2^{(q+1)/2}, 2^{q+1}), \quad (8)$$
$$r_0(y) = y^2, \ r_1(y) = y^2/2^{q+1}.$$

Then the final solution can be expressed as

$$\log_2 x = (q+1)f(y) \mp p. \quad (9)$$

For different values of $p$ and $q$, the main difference in quantum circuit design is the multiplication and addition factors. We take the case of $p=0$ and $q=1$ as the example to show the circuit design. In such case, Eq. (8) turns to

$$f(x) = (\log_2 x)/2, \ I = [1, 4),$$
$$D_0 = [1, 2), D_1 = [2, 4), \quad (10)$$
$$r_0(x) = x^2, \ r_1(x) = (x/2)^2.$$

Substitute these elements into Eq. (4), then we have

$$w_i = \begin{cases} 0, a_i < 2 \\ 1, a_i \geq 2 \end{cases}; \ a_0 = x, a_{i+1} = \begin{cases} a_i^2, a_i < 2 \\ a_i^2/2^2, a_i \geq 2 \end{cases}. \quad (11)$$

Fig. 2 shows the quantum circuit to implement logarithmic function based on Eq. (11). The circuit generally consists of ($n$-1) log-modules and each log module outputs one bit of the solution. The dominant cost of the present algorithm results from the square operation in each log module. Suppose that the number of qubits and quantum operations used in the circuit for square operation are $O(m)$ and $O(m^2)$ respectively, then the complexity of solving the logarithmic function by the qFBE method is ($n$-1)$O(m)$ in qubits and ($n$-1)$O(m^2)$ in quantum operations.

The log module executes through two steps as shown in the figure.

1st. Obtain $w_i$ based on the value of $a_i$. For the present case with $p=0$ and $q=1$, if $a_i$ belongs to $D_0$ = [1,2), then the highest bit of $a_i$ equals to 0; otherwise $a_i$ belongs to $D_1$ = [2,4) and highest bit is 1. So a CNOT gate is enough to determine the value of $w_i$ conditioned on the highest bit of $a_i$.

2nd. Compute $a_{i+1}$. The controlled Shift_R module is used to shift the binary string of of $a_i$ to right one bit, namely to obtain $a_i/2$ when $w_i = 1$. The Square module represents the square operation.

For the quantum circuits implementing square operation, Refs. [8,12-15] provide several kinds of options which base on different methodologies. Here we propose that the square circuit can be obtained by reversing the circuit for square root operation, and the square root circuit can be designed based on the non-restoring method. The non-restoring method is a classical digit-recurrence algorithm (being opposite to the functional iteration algorithm) which can be explored to calculate square root and reciprocal [26,27]. Ref. [21] provides feasible circuits design to implement square root and reciprocal based on the non-restoring method. Similar work can also be found in Refs. [17,18], which further optimizing the number of qubits and T-gate used in the circuit for the better performance in fault tolerant computing.

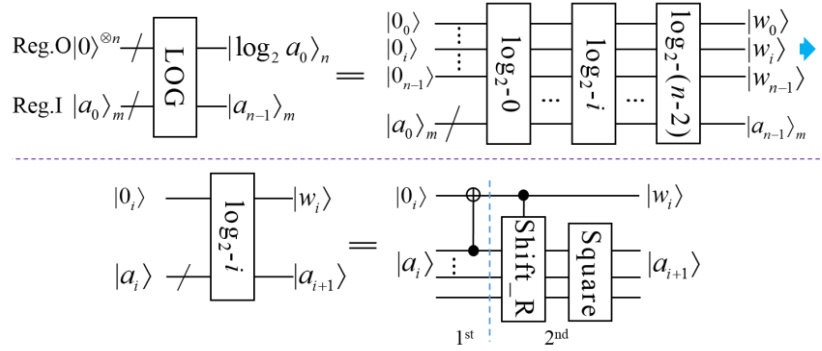

Fig. 2. The quantum circuit for implementing the logarithmic function based on Eq. (11). The upper half of the figure shows the overall structure of the circuit, and the lower half the design of the log module. The circuit mainly includes two registers, Reg. O and Reg. I, with the number of qubits of $n$ and $m$, respectively. Each log module contains one square operation. Note that the subscript $i$ and $m$ in $|a_i\rangle_m$ denote that $m$ qubits are used to store the $i^{th}$ intermediate value of $a$ in Eq. (4). $|0_i\rangle$ ($|w_i\rangle$) denotes the $i^{th}$ qubit of input (output) in Reg. O.

3.1.2 Arc cosine & arc sine functions

We will show the algorithm and circuit for the arc cosine function. The arc sine function can be evaluated easily using the angle relation with arc cosine function. The basic elements used to expand arc cosine function are as follows,

$$f(x) = \arccos(x)/\pi, I = (-1,1],$$
$$D_0 = (0,1], D_1 = (-1,0],$$
$$r_0(x) = 2x^2 - 1, r_1(x) = 1 - 2x^2. \quad (12)$$

Substitute these elements into Eq. (4), and then we have

$$w_i = \begin{cases} 0, a_i > 0 \\ 1, a_i \le 0 \end{cases}; \quad a_0 = x, a_{i+1} = \begin{cases} 2a_i^2 - 1, a_i > 0 \\ 1 - 2a_i^2, a_i \le 0 \end{cases}. \quad (13)$$

Fig. 3 shows the quantum circuit to implement arc cosine function based on Eq. (13). It generally consists of ($n$-1) arccos-modules and each arccos module outputs one bit of the solution, namely $w_i$. As it is in logarithm function, the dominant cost of the present algorithm results from the square operation in each arccos module. The complexity of solving the arccosine function by the qFBE method is ($n$-1)$O(m)$ in qubits and ($n$-1)$O(m^2)$ in operations, where $O(m)$ and $O(m^2)$ denotes the number of qubits and operations used in the circuit for square operation.

The arccos module executes through six steps as shown in the figure.

1st. Set the master control bit. According to Eq. (13), the values of $w_{i+1}$ up to $w_{n-1}$ are equal to 0 and the values of $a_{i+1}$ up to $a_{n-1}$ are equal to 1 when $a_i$ equals to 0. So we set Anc1 as the master control bit to determine the arithmetic operations.

2$^{nd}$. Obtain $w_i$. Since the value of $w_i$ can be determined by the sign of $a_i$, we can set the highest bit of $a_i$ be the sign bit. Then when $a_i$ is positive, the value of $w_i$ can be flipped. Note that when $a_i = 0$, $w_i$ is equal to 1, so it will be flipped twice to remain unchanged (See step 6).

3$^{rd}$. Obtain the complement of $a_i$. Since $a_i$ is a signed number, it should be converted to positive number to do operation when it is negative. We represent it by the form of binary complement, so the X gates and Adder* module do the operation of "invert and plus one". The superscript * in Adder* means the "plus one" is carried out at the lowest bit of $a_i$. So this step may be called the absolute value module.

4$^{th}$. Obtain $\pm a_{i+1}$. The calculation of $2a_i^2-1$ is done by the controlled square, Shift_L and subtraction operations. According to Eq. (13), $2a_i^2-1$ is just the case when $a_i > 0$, so the $\pm a_{i+1}$ is obtained. The controlled Shift_L module can shift the binary string of $a_i^2$ to left one bit, namely to obtain $2a_i^2$. The subtraction module minus one at the lowest bit of the integer part of $2a_i^2$.

5$^{th}$. Obtain $a_{i+1}$. It is same as that in third step.

6$^{th}$. Supplementary for the case of $a_i = 0$. Anc1 turns to be 1 when $a_i = 0$. Then the CNOT gate will flip the $w_i$ back to 1(see second step) and the TOFFOLI gate will make sure the values of $w_{i+1}$ to $w_{n-1}$ all equal to 0.

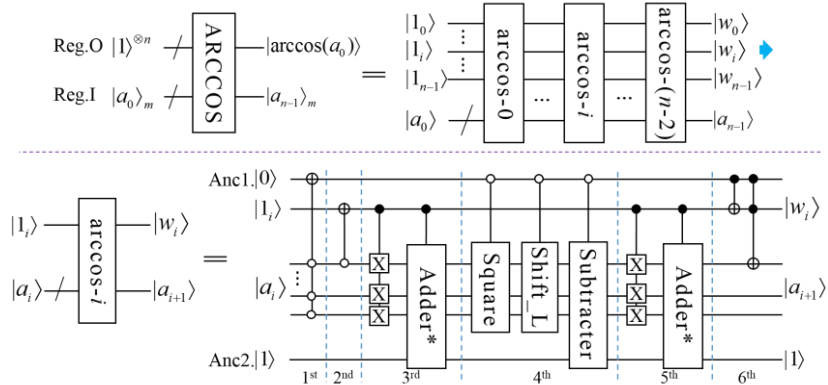

Fig. 3. The quantum circuit for implementing the arc cosine function based on Eq. (13). The upper half of the figure shows the overall structure of the circuit, and the lower half the design of the arccos module. Each arccos module contains one square operation. The X represents the NOT gate. The Adder* and Adder denotes different kind of addition operation.

3.1.3 Arc cotangent & arc tangent functions

We will show the algorithm and circuit for the arc cotangent function. The arc tangent function can be evaluated easily using the angle relation with arc cotangent function. The basic elements used to expand arc cotangent function are as follows,

$$f(x) = \arccot(x)/\pi, I = R \cup \{-\infty\},$$
$$D_0 = (0, \infty) \cup \{-\infty\}, D_1 = (-\infty, 0], \qquad (14)$$
$$r_0(x) = \frac{x^2-1}{2x}; r_1(x) = \frac{x^2-1}{2x}, r_1(0) = -\infty.$$

Substitute these elements into Eq. (4), and each bit of the solution can be obtained by

$$w_i = \begin{cases} 0, a_i \in (0,\infty) \cup \{-\infty\} \\ 1, a_i \in (-\infty, 0] \end{cases}; \quad a_0 = x, a_{i+1} = \begin{cases} \dfrac{a_i^2 - 1}{2a_i}, a_i \neq 0 \\ -\infty, a_i = 0 \end{cases}. \tag{15}$$

The quantum circuit to implement arc cotangent function is shown in Fig. 4. It generally consists of (n-1) arccot-modules and each arccot module outputs one bit of the solution. The dominant cost of the present algorithm results from the reciprocal operation in each arccot module. Typically, the complexity of the reciprocal circuit is $O(m)$ in qubits and $O(m^2)$ in quantum operations. So the complexity of solving the arccot function by the qFBE method is $(n-1)O(m)$ in qubits and $(n-1)O(m^2)$ in operations.

The arccot module is almost the same with arccos module. The difference is that the Square and Shift_L modules replace the Reciprocal and Shift_R modules, respectively. For the quantum circuits implementing reciprocal, the results in Refs. [17,21] are good choices, which are designed based on the non-restoring method.

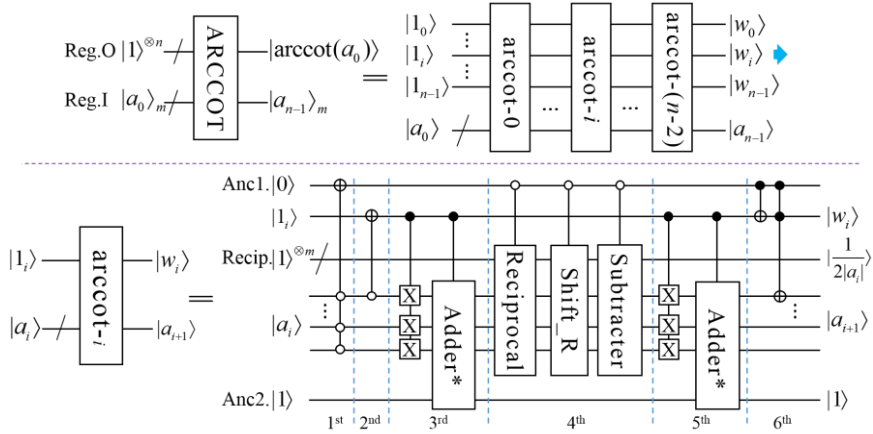

Fig. 4. The quantum circuit for implementing the arc cotangent function based on Eq. (15). The upper half of the figure shows the overall structure of the circuit, and the lower half the design of the arccot module. Each arccot module contains one reciprocal operation. The Adder* and Adder modules are the same as that in Fig. 3.

3.2 Quantum circuits for functions in Group 2
3.2.1 Exponential function

We will show the algorithm and circuit for the base-two exponential function. Other kinds of exponential functions can be evaluated by a further power operation. For example, the natural exponential function can be computed based on the base-two exponential function using the relation of $e^x = 2^{x \cdot \log_2 e}$.

Substitute Eq. (6) into Eq. (5), then we have the following iterative formula to approximate the base-two exponential function value,

$$a_0 = 1, a_{i+1} = \begin{cases} \sqrt{a_i}, v_i = 0 \\ \sqrt{2a_i}, v_i = 1 \end{cases}, \quad i = 0, 1, 2, \ldots, (n-1). \tag{16}$$

Note the inputs satisfy $x \in [0, 1)$ and can be expressed as $x = (0.v_{n-1} \cdots v_1 v_0)_2$ in binary form, and the final output $a_n$ is the solution. Let us take $f(0.7) = 2^{0.7}$ as the example to illustrate the iteration process. The value of $2^{0.7}$ is about 1.6245. First, we know 0.7 is 0.1011 in binary form with four digits. Then the iteration goes as follows.

1st. Since $v_0 = 1$ and $a_0 = 1$, we have $a_1 = (2\times1)^{0.5} \approx 1.4142$.
2nd. Since $v_1 = 1$, we have $a_2 \approx (2\times1.4142)^{0.5} \approx 1.6818$.
3rd. Since $v_2 = 0$, we have $a_3 \approx (1.6818)^{0.5} \approx 1.2968$.
4th. Since $v_3 = 1$, we have $a_4 \approx (2\times1.2968)^{0.5} \approx 1.6105$ …

It can be seen clearly the above computing process is the reverse of that for Logarithmic function. The intermediate values of $a_i$ bring truncation errors, which will accumulate gradually to the final step, i.e., the solution.

The quantum circuit to implement base-two exponential function is shown in Fig. 5. It uses $n$ exp-modules to approximate the function value and the final step output is the solution. The dominant cost of the circuit results from the square root operation in each exp module. The number of qubits and quantum operations used in the circuit for square operation are typically $O(m)$ and $O(m^2)$, so the complexity of solving the exponential function by the qFBE method is $nO(m)$ in qubits and $nO(m^2)$ in quantum operations.

The exp module executes through two steps as shown in the figure.

1st. Shift $a_i$ to left one bit to obtain $2a_i$ when $v_i = 1$.

2nd. Perform the square root operation to obtain $a_{i+1}$. As mentioned in the section of logarithmic function, the quantum square root circuits provided in Refs. [18,21] can be used, which are designed based on the classical non-restoring method.

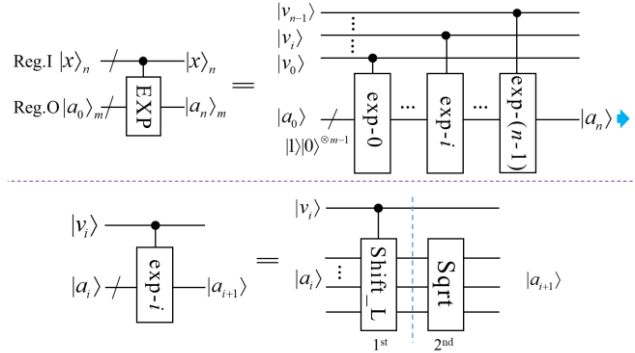

Fig. 5. The quantum circuit for implementing the base-two exponential function based on Eq. (16). The upper half of the figure shows the overall structure of the circuit, and the lower half the design of the exp module. Each exp module contains one square root operation. $|a_0\rangle$ is initialized as $|1\rangle|0\rangle^{\otimes m-1}$.

### 3.2.2 Cosine & sine functions

We will show the algorithm and circuit for the cosine function. The sine function can be evaluated easily using the angle relation with cosine function. Substitute Eq. (12) into Eq. (5), then we have the following iterative formula to approximate the cosine function value of $\cos(\pi x)$,

$$a_0 = 1; \ a_{i+1} = \begin{cases} \sqrt{(1+a_i)/2}, v_i = 0 \\ -\sqrt{(1-a_i)/2}, v_i = 1 \end{cases}, i = 0, 1, 2, \ldots, n-1. \tag{17}$$

The input $x$ is in the interval of $[0,1)$, and can be expressed as $x = (0.v_{n-1}\ldots v_1 v_0)_2$ in the binary form. The solution is $\cos(\pi x) = a_n$.

It is not friendly to design the corresponding quantum circuit based on Eq. (17) because $a_i$ is a signed number. For the input $x$ in the interval of $[0, 1/2]$, i.e., the corresponding function value is positive, Eq. (17) can be further written as

$$a_0 = 1; \ a_{i+1} = \begin{cases} \sqrt{(1+a_i)/2}, v_i = v_{i-1} \\ \sqrt{(1-a_i)/2}, v_i \neq v_{i-1} \end{cases}, i = 0, 1, 2 \ldots, (n-1), v_{-1} = 0. \tag{18}$$

The output of Eq. (18) is actually $a_n = |\cos(\pi x)|$, namely the absolute value of the solution. This is based on the observation that the sign of $a_{i+1}$ in Eq. (17) is determined by the value of $v_i$. The operation between 1 and $a_i$ under the square root is actually addition when $v_i = v_{i-1}$ and subtraction when $v_i \neq v_{i-1}$. The value of $a_i$ in Eq. (18) will always be positive. Finally we can obtain the real solution in the following way:

if $x \in [0, 1/2)$, namely $v_{n-1} = 0$, we have $\cos(\pi x) = a_n$;
if $x \in [1/2, 1)$, namely $v_{n-1} = 1$, we have $\cos(\pi x) = -a_n$.
The circuit is just a $v_{n-1}$-controlled absolute value module.

Fig. 6 shows the quantum circuit to implement cosine function based on Eq. (18). It uses $n$ cos-modules to approximate the function value and the final step output is the absolute value of the solution. The dominant cost of the circuit results from the square root operation in each cos module. The complexity of solving the cosine function by the qFBE method is $nO(m)$ in qubits and $nO(m^2)$ in quantum operations, where $O(m)$ and $O(m^2)$ are respectively the number of qubits and quantum operations used in the circuit for square operation.

The cos module executes through four steps as shown in the figure.

1st. Perform the addition operation of $1+a_i$ when $v_i = v_{i-1}$. It is just a normal addition operation.

2nd. Perform the subtraction operation of $1-a_i$ when $v_i \neq v_{i-1}$. Then if $1-a_i$ is not positive, the circuit of "invert and plus one" as that in Fig. 3 is used to obtain the complement.

3rd. Shift the value of $1 \pm a_i$ to right one bit to obtain $(1 \pm a_i)/2$.

4th. Perform the square root operation to obtain $a_{i+1}$.

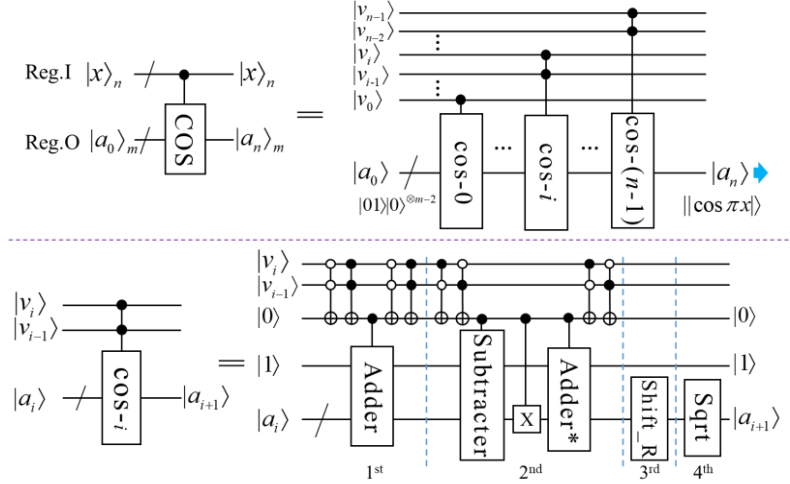

Fig. 6. The quantum circuit for implementing the cosine function based on Eq. (18). The upper half of the figure shows the overall structure of the circuit, and the lower half the design of the cos module. The final output is the absolute value of cosine function. Each cos module contains one square root operation. $|a_0\rangle$ is initialized as $|01\rangle|0\rangle^{\otimes m-2}$.

3.2.3 Cotangent & tangent functions

We will show the algorithm and circuit for the cotangent function. The tangent function can be evaluated easily using the angle relation with cotangent function. Substitute Eq. (14) into Eq. (5), then we have the following iterative formula to approximate the cotangent function value of $\cot(\pi x)$,

$$a_0 = +\infty;\ a_{i+1} = \begin{cases} a_i + \sqrt{a_i^2+1}, v_i = 0 \\ a_i - \sqrt{a_i^2+1}, v_i = 1 \end{cases}, i = 0,1,2,\ldots,(n-1). \quad (19)$$

The input $x$ is in the interval of (0, 1), and can be expressed as $x = (0.v_{n-1}\cdots v_1 v_0)_2$ in the binary form. The solution is $a_n = \cot(\pi x)$.

As it is for cosine function, Eq. (19) should be converted because $a_i$ is a signed number. The main difference is that the addition or subtraction operation is before or after square root operation. For the input $x$ in the interval of [0, 1/2], *i.e.*, the corresponding function value is positive, Eq. (19) can be further written as

$$a_0 = +\infty;\ a_{i+1} = \begin{cases} \sqrt{a_i^2+1}+a_i, v_i = v_{i-1} \\ \sqrt{a_i^2+1}-a_i, v_i \neq v_{i-1} \end{cases}, i = 0,1,2,\ldots,(n-1), v_{-1} = 0. \quad (20)$$

The output of Eq. (20) is actually $a_n = |\cot(\pi x)|$, namely the absolute value of the solution. Then we can obtain $\cot(\pi x)$ by a $v_{n-1}$-controlled absolute value module. The initial value of $a_0$ equals to $+\infty$ in Eq. (19) which is not friendly for the quantum circuit design. In fact, $a_0$ can be set as 1 due to the following observation: when from $v_0$ to $v_{i-1}$ are all 0, $a_i$ will remain; and once $v_i$ equals to 1, $a_{i+1}$ will turn to 0. So $a_0$ is not necessary to be equal to $+\infty$.

Fig. 7 shows the quantum circuit to implement cotangent function based on Eq. (20). Note that cot-0 module contains only step 3. So the COT circuit actually consists of ($n$-1) cot-modules and the final step output is the absolute value of the solution. The dominant cost of the circuit results from the square and square root operations in each cot module. The complexity of solving the cotangent function by the qFBE method is ($n$-1)$O(m)$ in qubits and ($n$-1)$O(m^2)$ in quantum operations, where $O(m)$ and $O(m^2)$ is respectively the number of qubits and quantum operations used in the circuit for square and square root operations.

The cot module executes through three steps as shown in the figure.

1st. Obtain $\sqrt{a_i^2+1}$ controlled by the qubit of Anc1. The Adder$^+$ and Square$^+$ modules are used to eliminate the garbage qubits.

2nd. Obtain $\sqrt{a_i^2+1} \pm a_i$. This step is similar to the first and second step in cosine function.

3rd. Change the master control qubit Anc1. As explained above, $a_0$ is initialized as 1. This step is used to judge whether $v_i = 1$ is satisfied. If satisfied, then the TOFFOLI gate turns $a_i = 1$ to $a_{i+1} = 0$. And the master control qubit Anc1 will be flipped to 0 by the next $a_{i+1}$-controlled NOT gates, which is the same as the first step in Fig. 3. The following arithmetic operations from $i+1$ to $n$-1 modules will be executed.

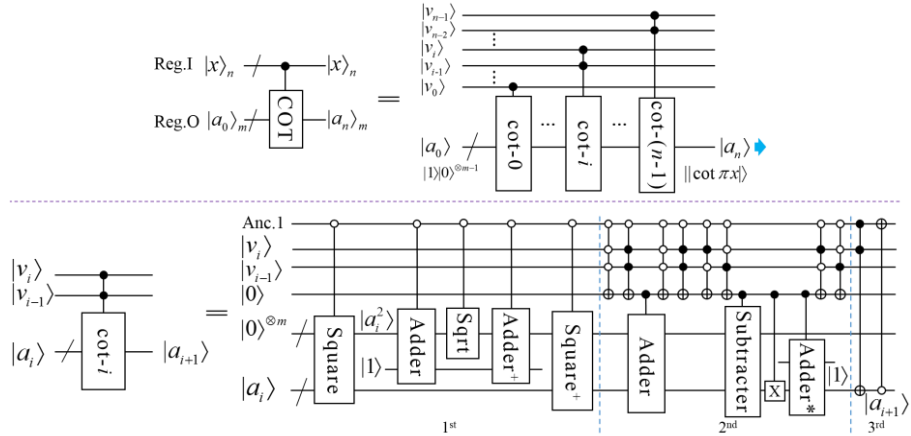

Fig. 7. The quantum circuit for implementing the cotangent function based on Eq. (20). The upper half of the figure shows the overall structure of the circuit, and the lower half the design of the cot module. The final output is the absolute value of cotangent function. Each cot module contains two square and one square root operations. $|a_0\rangle$ is initialized as $|1\rangle|0\rangle^{\otimes m-1}$.

## 4. Extension of qFBE to the qudit systems

Recently qudit systems get more and more attention due to its advantages over qubit system, like processing much larger information by less quantum hardware [28,29]. Here we would like to show that the present qFBE method can be extended into the qudit systems. Take the logarithmic function of Eq. (8) as the example, we can extend it into quantum ternary and quaternary systems, *etc*.

The basic elements used to expand logarithmic function value in a ternary form are as follows,

$$f(x) = (\log_2 x)/3, \ I = [1, 8),$$
$$D_0 = [1, 2), D_1 = [2, 4), D_2 = [4, 8), \quad (21)$$
$$r_0(x) = x^3, \ r_1(x) = x^3/8, \ r_2(x) = x^3/64.$$

The solution is $f(x)_3 = (0.w_0 w_1 \cdots w_{n-1})_3, w_i \in \{0, 1, 2\}$. Then the value of $\log_2 x$ can be obtained by a multiplication between $(10)_3$ and $f(x)_3$ in ternary systems. It is indeed a left shift operation of $f(x)_3$.

The basic elements used to expand logarithmic function value in a quaternary form are as follows,

$$f(x) = (\log_2 x)/2, \ I = [1, 4),$$
$$D_0 = [1, \sqrt{2}), D_1 = [\sqrt{2}, 2), D_2 = [2, 2\sqrt{2}), D_3 = [2\sqrt{2}, 4), \quad (22)$$
$$r_0(x) = x^4, r_1(x) = x^4/4, r_2(x) = x^4/4^2, r_3(x) = x^4/4^3.$$

The solution is $f(x)_4 = (0.w_0 w_1 \cdots w_{n-1})_4, w_i \in \{0, 1, 2, 3\}$. Then the value of $\log_2 x$ can be obtained by a multiplication between $2_4$ and $f(x)_4$ in quaternary systems. Same as Eq. (21), we can expand the interval $I$ to $[1, 16)$, and Eq. (22) turns to

$$f(x) = (\log_2 x)/4, \ I = [1, 16),$$
$$D_0 = [1, 2), D_1 = [2, 4), D_2 = [4, 8), D_3 = [8, 16), \quad (23)$$
$$r_0(x) = x^4, r_1(x) = x^4/4^2, r_2(x) = x^4/4^4, r_3(x) = x^4/4^6.$$

Then the value of $\log_2 x$ can be obtained by shifting the quaternary string of $f(x)_4$ left one quaternary bit.

## 5. Circuits demonstration on a quantum virtual system

We demonstrate our qFBE method for solving the transcendental functions on a quantum virtual computing system installed on the Sunway TaihuLight supercomputer of the National Supercomputing Center in Wuxi, China. It has three operation modes, which are Full Amplitude mode, Partial Amplitude mode and Single Amplitude mode. The codes running on the virtual system are written using the quantum assembly language of QRunes developed by the Origin Quantum Computing Company [30]. More information about this system can be found in Ref. [31].

The circuits for the logarithmic, arc cosine, arc cotangent, exponential and cosine functions are demonstrated individually. The LOG, ARCCOS, ARCCOT, EXP and COS circuits evaluate the functions of $\log_2 x$, $\arccos(x)/\pi$, $\text{arccot}(x)/\pi$, $2^x$ and $\cos(\pi x)$, respectively. We did not execute the cotangent circuit on the virtual system because of the large number of qubits and gates. The configurations of implementing the circuits on the quantum virtual system are listed in Tab. 1. Source codes of all circuits are provided as the supplementary material of the present paper, which can be found in Ref. [32].

Tab. 1. The configurations of executing our circuits on quantum virtual system

|  | LOG | ARCCOS | ARCCOT | EXP | COS |
|---|---|---|---|---|---|
| Qubits | 40 | 30 | 24 | 32 | 29 |
| Basic Gates [a] | 1260 | 770 | 1000 | 820 | 730 |
| Least Nodes | 4096 | 4 | 1 | 16 | 2 |
| Operation Mode | Full Amp. | Full Amp. | Full Amp. | Full Amp. | Full Amp. |
| Run time (min) | 23 | 13 | 4 | 14 | 14 |

[a] They include the *H*, *X*, *C-NOT*, *SWAP* and *TOFFOLI* gates.

The running results of our circuits are listed in Tab. 2. The input *x* and corresponding solutions are encoded as the binary string of the input and output quantum states, respectively. As can be seen from the results, each bit of the binary string of the solution is exact (except the lower bits of the solution of EXP when the input is .11) and the error is determined by the number of qubits, see appendix. The running results verify the effectiveness of our algorithms.

Tab. 2. The results of our circuits demonstrated on quantum virtual system

|  | Input States | Output States |
|---|---|---|
| LOG | $|01.00\rangle, |10.00\rangle$ | $|0.000\rangle, |1.000\rangle$ |
| ARCCOS | $|00.00\rangle, |00.10\rangle, |01.00\rangle, |11.10\rangle$ | $|.10\rangle, |.01\rangle, |.00\rangle, |.10\rangle$ |
| ARCCOT | $|01.00\rangle$ | $|.01\rangle$ |
| EXP | $|.00\rangle, |.01\rangle, |.10\rangle, |.11\rangle$ | $|01.00\rangle, |01.01\rangle$ |
| COS | $|00\rangle, |01\rangle, |10\rangle, |11\rangle$ | $|01.000\rangle, |00.101\rangle, |00.000\rangle, |11.011\rangle$ |

## 6. Conclusions

In the present work, we develop the qFBE (quantum Function-value Binary Expansion) method to evaluate the transcendental functions, including the logarithmic, exponential, arc cosine (arc sine), cosine (sine), cotangent (tangent) and arc cotangent (arc tangent) functions. This method transform the calculation of transcendental functions to the computation of algebraic functions. Actually, the invoked algebraic functions are very simple, including the addition, square, square root and shift left or right operations. A vast amount of literature provides quantum circuits for such functions. Furthermore, the qFBE method has the potential to be extended into the qudit systems.

In general, the complexity of solving the transcendental functions by the qFBE method are $O(n^3)$. The qFBE makes the circuit design modular and compact. We demonstrate the circuits on a quantum virtual computing system installed on the Sunway TaihuLight supercomputer. For the further work, we will first optimize the basic modules, *e.g.*, adder, square and square root, to reduce the complexity of the circuits, and then utilize qFBE to solve practical quantum problems.

**Acknowledgements**

We are very grateful to the National Supercomputing Center in Wuxi for the great computing resource. We would also like to thank the technical team from the Origin Quantum Computing Technology co., LTD in Hefei for the professional services on quantum virtual computation. The present work is financially supported by the National Natural Science Foundation of China (Grant No. 61575180, 61701464, 11475160) and the Pilot National Laboratory for Marine Science and Technology (Qingdao).

**Appendix**

In the appendix, we perform error analysis of the solution calculated by the qFBE method. We want to determine how many digits of the binary solution are exact, and then use it to control the error propagation of the algorithm.

For the logarithmic and inverse trigonometric functions, namely the functions in Group 1, each bit of the solution $w_i$ is determined by the fact that which sub-interval, $D_0$ or $D_1$, the $a_i$ belongs to. The solution will be an exact one except the truncation error when $a_i$ belongs to the corresponding $D_j$ correctly. So the solution error should not be sensitive to the accumulation error of the intermediate $a_i$. On the other hand, for the exponential and trigonometric functions, namely the functions in Group 2, the truncation errors of the intermediate $a_i$ accumulate gradually to the final step. That is, the error from every intermediate $a_i$ has directly impact on the solution.

In the following analysis, we assume that the fractional part of $a_i$ occupies $q$ qubits. We employ electronic design software to assist our error analysis, for the accumulation process of truncation errors is complicated in some cases. And it is reasonable to believe that the actual upper bounds are much lower than that analyzed in theory.

A.1 Logarithmic function

According to Eq. (11), the highest two bits of $a_i$ hold the integer part and the lower $m$-2 bits the fractional part, *i.e.*, $q = m-2$. In addition, we can observe the fact that when the input $x$ near the dividing point, the accumulated error of $a_i$ is most likely induce the flip of $w_i$. The largest error would be induced when $x = a_0 = 2+2^{-q/2+1}-2^{-q}$. We have not yet find out a proper theoretical formula to determine the error bound in the worst case. We turn to the numerical tests to simulate error accumulation by the Matlab software. The variable precision arithmetic (vpa) function is utilized to control the truncation error of each intermediate $a_i$. The numerical results show that if the number of qubits used to store $a_i$ is equal to that of output, namely $n = m$ as shown in Fig. 2, all the $m$ bits of the output can be guaranteed to the exact.

A.2 Arc cosine function

According to Eq. (13), the highest qubit of $a_i$ is the sign bit, and the second bit holds the integer part. So the lower $m$-2 bits represent the fractional part, i.e., $q = m$-2. As it is in the logarithmic function, when the input $x$ near the dividing point, the accumulated error of $a_i$ will induce larger error in the solution. The largest error in solution $f(x)$ is induced when the input is $x = a_0 = 2^{-q/2}(1-2^{-q/2})$. And the actual solution obtained by the circuit shown in Fig. 3 is $\hat{y}(a_0) = 2^{-1} - 2^{-q}$.

The derivative of function $y = \arccos(x)/\pi$ is $-1/(\pi\sqrt{1-x^2})$, which is decreasing in [0,1). So in the interval of [0,$a_0$], $y$ satisfies

$$|\frac{\Delta y}{\Delta x}| \leq |y'(a_0)| = \frac{1}{\pi\sqrt{1-a_0^2}} = \frac{1}{\pi\sqrt{1-2^{-q}(1-2^{-q/2})^2}} < \frac{1}{\pi(1-2^{-q})}. \quad (A.2.1)$$

When $\Delta x = a_0$, the equation turns to

$$|\Delta y| = \frac{1}{2} - y(a_0) < \frac{2^{-q/2}(1-2^{-q/2})}{\pi(1-2^{-q})} < 2^{-q/2-1}. \quad (A.2.2)$$

This is the lower bound of the solution. And the upper bound of error in actual solution satisfies

$$\hat{y}(a_0) - y(a_0) < (2^{-1} - 2^{-q}) - (2^{-1} - 2^{-q/2-1}) = 2^{-q/2-1} - 2^{-q}. \quad (A.2.3)$$

So we can obtain an exact solution of $n$ bits (see Eq. (7)) when $n \leq m/2+1$ in the worst case. See Fig. 3 for the meaning of $m$ and $n$.

The calculation of $a_{i+1}$ in Eq. (13) can be converted to the following equation to reduce the error slightly,

$$a_{i+1} = \begin{cases} (2a_i)^2/2 - 1, a_i > 0 \\ 1 - (2a_i)^2/2, a_i \leq 0 \end{cases}, \quad (A.2.4)$$

where the operation complexity remains the same.

A.3 Arc cotangent function

We again use the numerical method to simulate the effect of error accumulation in Matlab. The simulation results show that if the number of qubits used to store intermediate $a_i$ is equal to that of output, namely $n = m$ as shown in Fig. 4, all the $m$ bits of the output can be guaranteed to the exact.

A.4 Exponential function

According to Eq. (16), the highest bit of $a_i$ holds the integer part and the lower $m$-1 bits the fractional part, i.e., $q = m$-1. Based on the relation of $\sqrt{1+2^{-q}} < 1+2^{-q-1}$, we have the following inequality when $\hat{a}_i$ is not less than $\sqrt{2}$,

$$\sqrt{\hat{a}_i + 2^{-q}} = \sqrt{\hat{a}_i}\sqrt{1+2^{-q-\log_2 \hat{a}_i}} < \sqrt{\hat{a}_i}(1+2^{-q-1-\log_2 \hat{a}_i})$$
$$= \sqrt{\hat{a}_i} + 2^{-q-1-\frac{1}{2}\log_2 \hat{a}_i} < \sqrt{\hat{a}_i} + 2^{-q-5/4}. \quad (A.4.1)$$

When the input $x$ equals to $(0.\underbrace{1\cdots 1}_{n})_2$, the accumulation of errors can be induced as follows,

$$\begin{cases} a_1 - \hat{a}_1 = \sqrt{2a_0} - \sqrt{2\hat{a}_0} + \varepsilon_1 = \varepsilon_1 < 2^{-q} \\ a_2 - \hat{a}_2 = \sqrt{2a_1} - \sqrt{2\hat{a}_1} + \varepsilon_2 < 2^{-q-3/4} + 2^{-q} < 2^{-q+1} \\ a_3 - \hat{a}_3 = \sqrt{2a_2} - \sqrt{2\hat{a}_2} + \varepsilon_3 < 2^{-q+1/4} + 2^{-q} < 2^{-q+2} \\ a_4 - \hat{a}_4 = \sqrt{2a_3} - \sqrt{2\hat{a}_3} + \varepsilon_4 < 2^{-q+5/4} + 2^{-q} < 2^{-q+2} \\ \cdots \\ a_n - \hat{a}_n = \sqrt{2a_{n-1}} - \sqrt{2\hat{a}_{n-1}} + \varepsilon_n < 2^{-q+2}. \end{cases} \quad (A.4.2)$$

At the worst, the high $q$-2 bits of the fractional part of $a_n$ guarantee no error. So the high $m$-2 bits of the solution involve no error.

A.5 Cosine function

According to Eq. (18), the highest two bits of $a_i$ hold the integer part and the low $m$-2 bits the fractional part, i.e., $q = m-2$. The maximum error-accumulation occurs at solutions of $\cos(1/2 \pm 1/2^n)\pi$. The case of $x = (0.0\underbrace{11\cdots 1}_{n-1})_2$ is chosen to evaluate the upper bound of the error. Eq. (18) can be expressed as

$$y(x) = \begin{cases} \sqrt{(1+x)/2}, v_i = v_{i-1} \\ \sqrt{(1-x)/2}, v_i \neq v_{i-1} \end{cases}, x \in [0,1], y \in [0,1]. \quad (A.5.1)$$

The derivative of the first iterative function is decreasing and belongs to the interval of $[1/4, \sqrt{2}/4]$, while the second is also decreasing and belongs to the interval of $[-\sqrt{2}/4, -\infty)$. The iteration process is as follows,

$$\begin{cases} \hat{a}_1 = y(a_0) + 0 \\ \hat{a}_2 = y(a_1) + \varepsilon_2 \\ \hat{a}_3 = y(\hat{a}_2) + \varepsilon_3 \\ \cdots \\ \hat{a}_n = y(\hat{a}_{n-1}) + \varepsilon_n \end{cases}, \quad (A.5.2)$$

where $\varepsilon_i$ represents the truncation error of each iteration step. The error of the first step is zero. The next $n$-2 steps satisfy $v_i = v_{i-1}$, so the first iterative function is calculated $n$-1 times. The last step calculates the second iterative function. Using the fact that the derivative of $\sqrt{(1+x)/2}$ is no larger than $\sqrt{2}/4$, the accumulation of errors can be

expressed as

$$\begin{aligned}
|a_n - \hat{a}_n| &= |y(a_{n-1}) - y(\hat{a}_{n-1}) - \varepsilon_n| \\
&< \delta |a_{n-1} - \hat{a}_{n-1}| + |\varepsilon_n| \\
&< \frac{\sqrt{2}}{4}\delta |a_{n-2} - \hat{a}_{n-2}| + \delta|\varepsilon_{n-1}| + |\varepsilon_n| \\
&\cdots \\
&< \delta((\frac{\sqrt{2}}{4})^{n-3}|\varepsilon_2| + \ldots + \frac{\sqrt{2}}{4}|\varepsilon_{n-2}| + |\varepsilon_{n-1}|) + |\varepsilon_n| \\
&< [\frac{1-(\sqrt{2}/4)^{n-2}}{1-\sqrt{2}/4}\delta + 1]2^{-q} \\
&< [4\delta + 1]2^{-q},
\end{aligned} \tag{A.5.3}$$

where $\delta$ is the derivative of $\sqrt{(1-x)/2}$ and satisfies $\delta = |y'_n| \in [\sqrt{2}/4, +\infty)$.

Now we evaluate the upper bound of $\delta$. Based on the relation of $\cos(x) < 1 - x^2/4$ where $x$ belongs to $(0,1)$, the output of the $n$-1$^{th}$ iteration satisfies

$$\hat{a}_{n-1} < a_{n-1} = \cos\frac{\pi}{2^{n-1}} < 1 - \frac{1}{4}(\frac{\pi}{2^{n-1}})^2 < 1 - 2^{-2n+3}, \tag{A.5.4}$$

so the absolute value of derivative of $y_n$ is

$$|y'_n| = \frac{1}{\sqrt{2}}\frac{1}{\sqrt{1-\hat{a}_{n-1}}} < \frac{1}{\sqrt{2}}\frac{1}{\sqrt{2^{-2n+3}}} = \frac{1}{\sqrt{2^{-2n+4}}} = 2^{n-2}, \tag{A.5.5}$$

substitute Eq. (A.5.5) into Eq. (A.5.3), and we obtain the upper bound

$$|a_n - \hat{a}_n| < (4 \cdot 2^{n-2} + 1)2^{-q} = (2^n + 1)2^{-q} = (2^n + 1)2^{-m+2}. \tag{A.5.6}$$

At the worst, the high $q$-$n$ bits of the fractional part of $a_n$ guarantee no error. So the high $m$-$n$ bits of the solution involve no error.

A.6 Cotangent function

We utilize Matlab to simulate numerically the effect of error accumulation. The results show that if the number of qubits used to store intermediate $a_i$ is equal to that of output, namely $n = m$ as shown in Fig. 7, almost all of the $m$ bits of the output can be guaranteed to the exact.